\documentclass[final,1p,times]{elsarticle}
\usepackage{amssymb}

\usepackage{amsmath}
\usepackage{amsthm}

\usepackage{tabularx}
\usepackage{booktabs}
\usepackage{multirow}

\usepackage{tikz}
\usepackage[font={sf,small}]{caption}
\usetikzlibrary{arrows.meta, positioning,calc}

\journal{Journal of Sound and Vibration}

\begin{document}

\begin{frontmatter}

%% Title, authors and addresses

%% use the tnoteref command within \title for footnotes;
%% use the tnotetext command for theassociated footnote;
%% use the fnref command within \author or \affiliation for footnotes;
%% use the fntext command for theassociated footnote;
%% use the corref command within \author for corresponding author footnotes;
%% use the cortext command for theassociated footnote;
%% use the ead command for the email address,
%% and the form \ead[url] for the home page:
%% \title{Title\tnoteref{label1}}
%% \tnotetext[label1]{}
%% \author{Name\corref{cor1}\fnref{label2}}
%% \ead{email address}
%% \ead[url]{home page}
%% \fntext[label2]{}
%% \cortext[cor1]{}
%% \affiliation{organization={},
%%             addressline={},
%%             city={},
%%             postcode={},
%%             state={},
%%             country={}}
%% \fntext[label3]{}

\title{Can a Building Work as a Reservoir: Footstep Localization with Embedded Accelerometer Networks}

%% use optional labels to link authors explicitly to addresses:
\author[label1]{Jun Wang}
\author[label2]{Rodrigo sarlo} %% Author name
\author[label1]{Suyi Li}

\affiliation[label1]{organization={Virginia Tech},
            addressline={Goodwin Hall},
            city={Blacksburg},
            postcode={24061},
            state={VA},
            country={USA}}

\affiliation[label2]{organization={Virginia Tech},
            addressline={Patton Hall},
            city={Blacksburg},
            postcode={24061},
            state={VA},
            country={USA}}

%% Abstract
\begin{abstract}
%% Text of abstract

Using floor vibrations to accurately predict occupants' footstep locations is essential for smart building operation and privacy-preserving indoor sensing.  However, existing approaches are dominated by either physics-based models that rely on simplified wave propagation assumptions and careful calibration, or data-driven methods that require large labeled datasets and often lack robustness to subject and environmental variability. This work introduces a new approach by treating an instrumented building floor as a physical reservoir computer, whose intrinsic structural dynamics can perform nonlinear spatio–temporal computation and information extraction directly. Specifically, foot strike-induced floor vibrations recorded by a distributed accelerometer network are processed using a lightweight physical reservoir computing (PRC) pipeline consisting of short waveform extraction, root-mean-square (RMS) normalization, principal component analysis (PCA), and a weighted linear readout. 
Results of this study, involving 2 participants and 12 accelerometers, showed that RMS normalization and PCA projection successfully extracted occupant-invariant features from the floor vibration waveform data, so that a single linear readout can predict foot strike location across repeated traversals and across multiple participants. Sub-meter accuracy is achieved along the hallway direction with moderate sensing coverage, while cross-participant tests achieved meter-scale accuracy without subject-specific recalibration or retraining. These findings demonstrate that building-scale structures can function as capable physical reservoir computers for intelligent monitoring.

\end{abstract}

%% Keywords

\end{frontmatter}

%% Add \usepackage{lineno} before \begin{document} and uncomment 
%% following line to enable line numbers
%% \linenumbers

%% main text
%%

%% Use \section commands to start a section
\section{Introduction --- Use the Building as a Physical Reservoir to Compute Footstep Location}

Accurately locating human footsteps inside a building is an important capability for smart civil infrastructure because it can enable numerous applications such as occupant tracking, energy-aware HVAC control, security monitoring, and emergency response~\cite{Mirshekari2018Occupant, Mirshekari2019PRCGait, Mirshekari2021Obstruction-invariant}. Currently, indoor localization strategies broadly include camera-based systems~\cite{Li2020Smart, Dong2022GaitVibe+:}, wearable-based sensing~\cite{Chen2024Motion}, and vibration-based approaches that exploit floor responses~\cite{alajlouni2020passive, Poston2017Indoor, Drira2022A}. Among these, camera-based localization could raise privacy concerns and typically requires continuous visual coverage and line-of-sight. Meanwhile, wearable-based methods are constrained by user compliance, device availability, and reliable communication, limiting their scalability for passive, building-wide deployment. In contrast, vibration-based localization only uses floor-mounted accelerometers, so it is inherently privacy-preserving, requires no line-of-sight, and can operate robustly in cluttered indoor environments even when sensors are sparsely deployed. When a human footstep impacts the floor, it excites dispersive bending and shear waves, with unique spatio-temporal patterns depending on the excitation location, walking dynamics, and the structural properties of the floor system. As a result, when sampled by a spatially distributed accelerometer network, these vibration fields can reveal footstep locations with sub-meter accuracy in realistic building environments \cite{Alajlouni2021MLE, Tarazaga2018Energy}.

\subsection{State of the Art in Vibration-Based Footstep Localization}

Existing vibration-based footstep localization methods can be broadly categorized into two paradigms: physics-based modeling approaches and data-driven approaches. Physics-based methods explicitly model the propagation of footstep-induced vibrations using structural dynamics, Lamb wave theory, and finite element representations. Representative techniques include time-difference-of-arrival (TDoA)~\cite{Appelle2024TDOA, Yang2021Survey}, energy-based localization~\cite{alajlouni2020passive,alajlouni2022maximum}, and error-domain model falsification (EDMF)~\cite{Drira2021EDMF, Ambarkutuk2024Modeling}, which infer source locations by accounting for dispersion, attenuation, and structural heterogeneity. These methods typically achieve average localization errors of approximately 0.4–0.8~m in real buildings and perform well in complex or obstructed environments. However, their accuracy depends on the availability of reliable structural models and appropriate parameter calibration, and it may be affected by modeling uncertainty or changes in boundary conditions.

In contrast, data-driven and machine-learning approaches, like the support vector machines (SVM)~\cite{Yu2023A} and convolutional neural networks(CNN)~\cite{Yu2020Deep, Dong2023Characterizing}, learn direct mappings from vibration features or raw waveforms to footstep locations. By exploiting large datasets, these methods excel at capturing nonlinear relationships, handling overlapping signals, and adapting to new environments, all leading to improved generalization across different floor types and occupant gaits~\cite{Mirshekari2020Step-Level, Mirshekari2019Physics-Guided}. Despite their accuracy, data-driven approaches typically require a substantial amount of labeled data, incur higher computational cost, and offer limited physical interpretability---these drawbacks can hinder real-time deployment and robustness under changing conditions. Therefore, recent studies are increasingly exploring hybrid strategies that combine physical insight with learning-based adaptability~\cite{Ambarkutuk2024Modeling, Imbiriba2024Augmented}, though many still rely on explicit models or large labeled datasets. 

\subsection{Physical Reservoir Computing for Information Perception}

In this study, we show that \textbf{P}hysical \textbf{R}eservoir \textbf{C}omputing (\textbf{PRC}) provides a unique pathway to footstep localization by bridging the physics-based and data-driven approaches: embedding physical dynamics directly into the learning process. \textit{In the big picture, a physical system with nonlinear dynamic properties (e.g., a real building) can project input signals (footsteps) into a high-dimensional state space (accelerometers' readings) with rich physical information. Therefore, by performing a simple weighted linear summation on this state space via a set of trained ``readout weights'', one can extract this information about the input (e.g., footstep location) }(Fig. \ref{fig:building_reservoir})~\cite{Nakajima2020Physical, Tanaka2019}. PRC has emerged as a powerful framework for information perception across different physical substrates. For example, photonic and optoelectronic reservoirs have enabled ultrafast, low-power processing for vision and olfactory sensing~\cite{Leng2024A, Huang2024Fully, Lao2022Ultralow‐Power}. In parallel, mechanical reservoirs---such as origami-inspired structures~\cite{bhovad2021physical}, soft robots~\cite{Wang2025Re-purposing, Wang2024Proprioceptive}, and compliant materials~\cite{cucchi2022hands}---have demonstrated much success in sensing and perception tasks, where their nonlinear deformation or vibration can naturally encode environmental information. For example, our previous study demonstrated that a simple, paper-based origami structure with load-bearing capacity can function as a physical reservoir~\cite{Wang2023Building}, capable of predicting payload weight and position as well as the input frequency by leveraging its high-dimensional dynamics. Across information-processing tasks, PRC systems often achieve accuracy comparable to or exceeding that of conventional machine-learning models, while offering advantages in computational efficiency, hardware simplicity, and real-time operation.

To date, however, PRC has been explored primarily in micro and meso-scale systems. Its application to large building-scale structures---and specifically to spatial inference problems such as footstep localization---remains largely unexplored. Yet civil structures inherently exhibit rich and nonlinear dynamic vibrations that can be sampled by embedded accelerometer networks. These characteristics suggest a compelling opportunity for PRC at the architectural scale.

\subsection{Contributions and Paper Organization}

Therefore, the central question addressed in this paper is: \emph{Can a building floor system---instrumented with embedded sensors---function as a physical reservoir that can transform footstep-induced vibrations into linearly decodable and real-time representations of footstep locations?}  Our answers to this question constitute the three major contributions of this paper
%
%This work proposes a vibration-based PRC pipeline with \textit{minimal} data processing, involving short waveform windows with Rooted-Mean-Square~(RMS) normalization, Principal Component Analysis~(PCA) projection, and finally a linear readout (Fig. \ref{fig:building_reservoir}). Such a setup can stabilize subject-dependent amplitude variability while preserving the dominant, location-dependent vibration modes.
%
\begin{itemize}
    \item \emph{Building-scale PRC for localization:} We experimentally show that a reinforced-concrete corridor floor with a sparsely embedded accelerometer network forms a high-dimensional spatio-temporal reservoir that encodes footstep location in its vibration response.
    \item \emph{Minimal vibration-based PRC pipeline:} We introduce an efficient, waveform-driven architecture for the vibration-based PRC using short sampling windows, RMS normalization, PCA projection, and a linear readout. Such a pipeline does not require complex data processing, structural modeling, or recurrent training.
    \item \emph{Data-efficient and cross-occupant performance:} We demonstrate accurate localization with few training traversal data and show that a readout trained on one subject can generalize to an unseen subject with sub-meter accuracy.
\end{itemize}

The remainder of the paper is organized as follows. Section~2 presents the experimental setup and evaluation protocols. Section~3 details the proposed PRC pipeline. Section~4 reports single and cross-occupant localization results, including training and sensor-reduction studies. Section 5 discusses physical observability and its implications for building-scale PRC. Section~6 concludes the paper and outlines future directions.

\section{Methods --- Conceptual Framework, Experimental Setup, and Sensory Data Processing}

This section first outlines the physical intuition behind the real-time foot strike location prediction using vibration-based physical reservoir computing. Then, we describe the building-scale experimental setup and the training/testing protocols.  Details of the event-detection algorithm, reservoir-state formalization, linear readout training, and error metric calculations are provided in the following subsections.

\subsection{Conceptual Framework: Using the Instrumented Building as a Physical Reservoir}

\begin{figure*}[b!]
    \centering
    \includegraphics[width=0.99\textwidth]{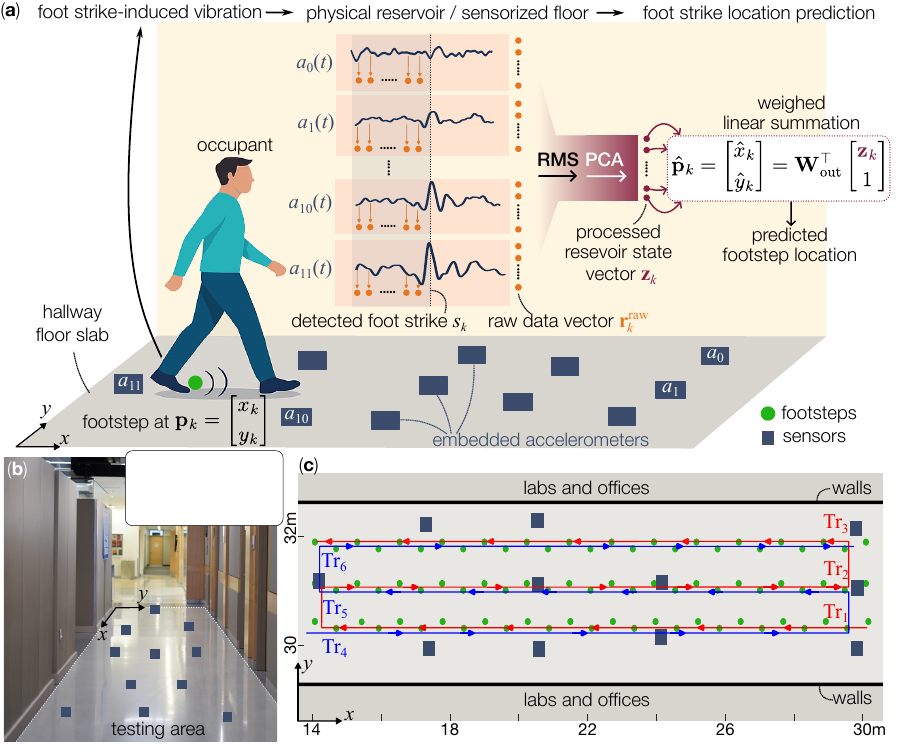}
    \caption{
    \textbf{Using an instrumented building as a physical reservoir computer for footstep localization.}
    (\textbf{a}) A walking human subject injects a mechanical impulse into the hallway floor at location $\mathbf{p}_k=(x_k,y_k)$, exciting dispersive vibration fields. A network of implanted floor accelerometers samples these fields, producing a high-dimensional signal stream $a_j(t)$. These signals will be extracted into short waveform windows, normalized via RMS, and projected via PCA into ``compact'' reservoir state vectors $\mathbf{z}_k$. Therefore, a simple weighted linear summation of $\mathbf{z}_k$ with trained readout weights $\mathbf{W}_\text{out}^\top$ can yield an accurate estimation of footstep location $\hat{\mathbf{z}}_k=(\hat{x}_k,\hat{y}_k)$.
    (\textbf{b}) Photograph of the instrumented hallway.
    (\textbf{c}) Top-view the hallway layout, showing accelerometer locations, walking traversals ($\text{Tr}_i$), and the global coordinate system.
    }
    \label{fig:building_reservoir}
\end{figure*}

As the human foot strike injects a localized, impulsive excitation to the floor slab, it generates transient vibration fields that propagate through the hallway as dispersive wave packets. 
% Due to reflections from the structural boundaries and spatially varying propagation paths, this vibration field is highly nonlinear and rich in spatio-temporal features. 
Such a transformation from impulse input to dispersive wave-field is high-dimensional and nonlinear, motivating us to treat the instrumented hallway and its sensor network as a \emph{physical reservoir computer}. More specifically, 
%
% the building itself performs the nonlinear spatio-temporal expansion of the footstep excitation, while a simple readout layer materializes the learning. From a reservoir computing perspective, 
%
the sensorized hallway structure performs a physical transformation:
\begin{equation}
\text{foot strike impulse at position } \mathbf{p}_k=\left(x_k,y_k\right)\;\;\mapsto\;\;\left\{a_j(t)\right\}_{j=0}^{N_s-1},
\end{equation}
meaning a localized mechanical input at position $\mathbf{p}_k=\left(x_k,y_k\right)$ is transformed into a high-dimensional vibration response field. (In this study, index $_k$ refers to unique foot step strikes.) This transformation is governed entirely by the building’s geometry, material properties, and boundary conditions.
%
%, and requires no explicit physical modeling (hence no physical parameter identification or calibration).
%
A network of implanted accelerometers samples this field at different locations, producing a waveform dataset $a_j(t)$. (The index $_j$ refers to different sensors, Figure~\ref{fig:building_reservoir}a). These waveforms constitute the raw \textit{reservoir states} that implicitly encode the footstep's spatial information in the form of sensing signal time delay, phase relationships, and vibration modal content. \textit{The reservoir computing framework proposes that, with an appropriate data processing setup, a weighted linear summation of these reservoir states can directly extract the foot strike locations.}
%
% Simple preprocessing—RMS normalization followed by PCA—conditions these high-dimensional states into a compact latent representation, which is mapped directly to the estimated footstep location using a linear readout. Crucially, no recurrent dynamics, iterative optimization, or physics-based inversion is required: the nonlinear expansion is performed physically by the structure itself.

\subsection{Experiment Setup}

All walking experiments were conducted in a hallway located on the fourth floor of Goodwin Hall on Virginia Tech’s campus~\cite{alajlouni2022maximum}.
The test area is a straight, approximately 16\,m long section of a reinforced-concrete hallway floor (Figure~\ref{fig:building_reservoir}b,c). 
The global coordinate system $(x,y)$ is aligned with the hallway, with the $x$-axis orienting along the walking direction and the $y$-axis pointing across the hallway. Over the instrumented floor area, $x$ ranged from roughly 14 to 30m, while $y$ was confined to a narrow region around 31.2m, corresponding to the hallway's centerline. The testing area is bounded by structural walls and adjacent laboratory spaces, producing a realistic indoor vibration environment with strong reflections and modal interactions. 

Eleven piezoelectric accelerometers were permanently mounted to the underside of the floor slab. Each sensor operated over a frequency range of 2–10,000 Hz, with a nominal sensitivity of 1000 mV/g and a resolution of approximately $8\times10^{-4}\mathrm{m/s^2}$. 
%
%Their locations were measured using a laser rangefinder and are stored in $\mathbf{S}\in\mathbb{R}^{2\times 11}$, where each column 
% $\mathbf{S}_i=[x_i,y_i]^{\top}$ denoted sensor~$i$'s position in the plane view (Fig.~\ref{fig:building_reservoir}c). 
%
% The sensors span horizontally and vertically along the corridor, providing sensitivity to both longitudinal (x) and lateral (y) variations in footstep location. 
%
Acceleration data were recorded using EMX-4250 digital signal analyzer cards with built-in anti-aliasing filters and 24-bit ADCs. All accelerometer channels were sampled synchronously at a rate of 1024 Hz.

Two adult subjects (denoted as $\text{S}_1$ and $\text{S}_2$) participated in the experiments. Each subject traversed the instrumented hallway six times along predefined routes. We denote them as $\text{Tr}_1$ to $\text{Tr}_6$ correspondingly (Figure~\ref{fig:building_reservoir}c), yielding a total of twelve traversals across both subjects. In each traversal, the subject walked at a comfortable and self-selected pace until he passed the sensorized floor region, then he turned around and re-entered the hallway for the subsequent traversal. To ensure consistency, every heel-strike locations were marked on the floor with tape, so that the subjects needed to step on these markers as accurately as possible. These markers defined 81 foot strike locations arranged along three parallel walking lines spanning the hallway, with a nominal step length of 0.62m and a stance width of 0.15m. Note that, although the walking routes were predefined, individual foot strikes still varied naturally in terms of timing, impact strength, and lateral placement.

\subsection{Physical Reservoir Computing (PRC) Pipeline}
To accommodate the unique dynamics and computing task requirements in this study, we formulated a new physical reservoir computing pipeline by using short windows of vibration waveforms as the reservoir states. 
%
% This way, we can eliminate the internal recurrent dynamics while preserving the nonlinear spatio-temporal expansion provided by the physical structure. 
%
This way, learning is reduced to training a weighted linear summation of these reservoir states to recover the foot strike location $\mathbf{p}_k$. 
%
% This formulation retains conceptual simplicity, incurs negligible computational overhead, and remains fully compatible with real-time, per-footstep localization. 

\begin{figure*}[htbp]
\centering
\begin{tikzpicture}[
    font=\small,
    box/.style={draw, rectangle, rounded corners=2pt, align=center, minimum height=9mm, minimum width=22mm},
    smallbox/.style={draw, rectangle, rounded corners=2pt, align=center, minimum height=8mm, minimum width=10mm},
    arrow/.style={-{Latex[length=2mm,width=1.5mm]}, line width=0.6pt},
    dashedarrow/.style={-{Latex[length=2mm,width=1.5mm]}, dashed, line width=0.6pt}
]

% ---------- Row 1: Sensing and event formation ----------
\node[box] (raw) {raw data from\\accelerators\\$a_j(t)$};
\node[box, right=7mm of raw] (det) {foot strike\\detection\\$s_k$};
\node[box, right=7mm of det] (win) {waveform\\windowing\\$t_w$};
\node[box, right=7mm of win] (vec) {reservoir state\\ vectorization\\$\mathbf{r}_k^\text{raw}$};

\draw[arrow] (raw) -- (det);
\draw[arrow] (det) -- (win);
\draw[arrow] (win) -- (vec);

% ---------- Row 2: Reservoir state and localization ----------
\node[box, below=7mm of raw] (rms) {RMS\\normalization\\$\mathbf{r}_k$};
\node[box, below=7mm of det] (pca) {PCA\\projection\\$\mathbf{z}_k$};
\node[box, below=7mm of win] (lin) {weighted linear\\ summation\\$\hat{\mathbf{p}}_k=(\hat{x}_k,\hat{y}_k)$};
\node[box, right=7mm of lin] (kf) {optional\\Kalman filter\\$(\hat{x}_k^\text{KF},\hat{y}_k^\text{KF})$};

% Replace the old vec -> rms arrow with:
\coordinate (route1) at ($(vec.south)+(0,-3mm)$);
\coordinate (route2) at ($(raw.south)+(0,-3mm)$);
\draw[arrow] (vec.south) --(route1) -- (route2) -- (rms.north);

\draw[arrow] (rms) -- (pca);
\draw[arrow] (pca) -- (lin);
\draw[dashedarrow] (lin) -- (kf);

% ---------- Training-only side inputs ----------
\node[smallbox, below=7mm of pca] (trainpca) {principal components\\$\mathbf{V}_D$};
\draw[dashedarrow] (trainpca) -- (pca);

\node[smallbox, below=7mm of lin] (trainw) {readout weights\\$\mathbf{W}_{\text{out}}$};
\draw[dashedarrow] (trainw) -- (lin);

\end{tikzpicture}

\caption{Real-time PRC footstep localization pipeline.}
\label{fig:realtime_pipeline}
\end{figure*}
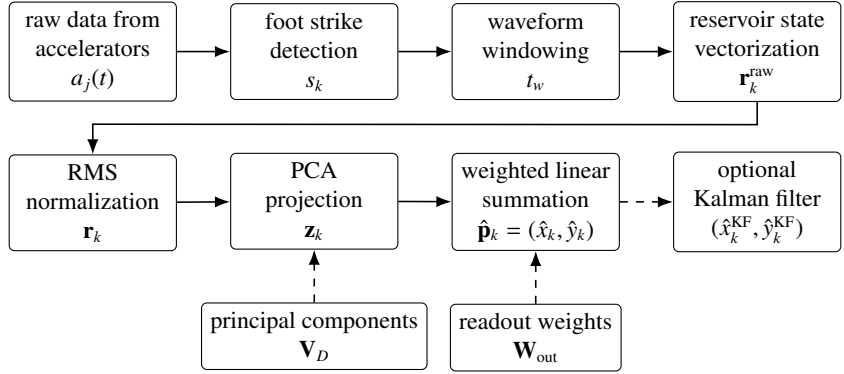

Figure~\ref{fig:realtime_pipeline} summarizes the PRC pipeline of this study. In real-time, we continuously stream and monitor the floor acceleration signals $a_j(t)$, and use a peak-picking algorithm to detect the occurrence of foot strikes. Once a footstep is detected at time stamp $s_k$, we extract a window of vibration waveform data from all sensors and vectorize these data into the raw reservoir state vectors $\mathbf{r}_k^{\mathrm{raw}}$. This window had a fixed length $t_w$ and position immediately before $s_k$ (Figure~\ref{fig:building_reservoir}a).
Then, we normalize and condition the raw data---with Root Mean Square (RMS) and Principal Component Analysis (PCA)---into a new set of compact reservoir state vectors $\mathbf{z}_k$. Here, the RMS normalization reduces the variability between subjects, and PCA projects the data into a lower-dimensional space containing the most relevant features.
Finally, we use a trained set of readout weights $\mathbf{W}_\text{out}$ to perform a weighted linear summation on $\mathbf{z}_k$, which could yield a prediction of the footstep locations $\hat{\mathbf{p}}_k=(\hat{x}_k,\hat{y}_k)$. Optionally, one can use a Kalman filter to refine the footstep predictions to smooth these output predictions.
%
% Each footstep is processed independently, enabling online localization with latency determined by the waveform window length. The detailed implementation of each processing stage, along with the rationale for each design choice, is described as follows.
%
In what follows, we detail this physical reservoir computing pipeline at every step. 

\subsubsection*{Footstep Event Detection}

The first step of the PRC pipeline is real-time detection of individual footstep strikes from the floor accelerometers' raw data.

\begin{figure*}[t!]
    \centering    
    \includegraphics[width=0.99\textwidth]{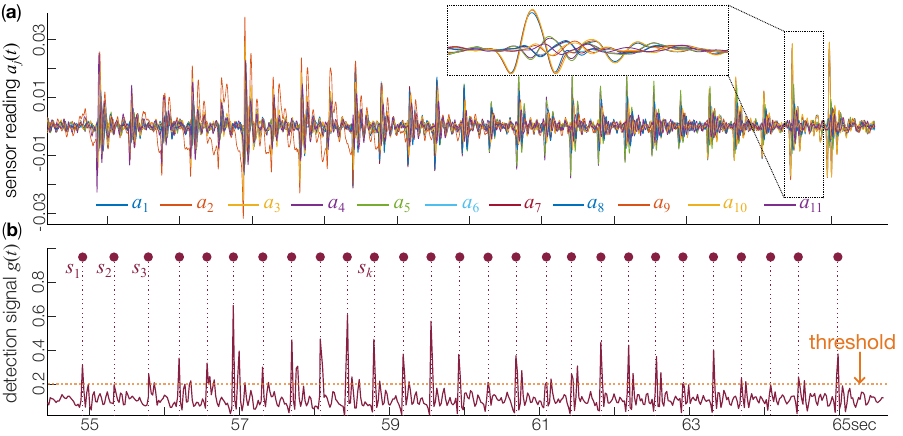}
    \caption{
    \textbf{Foot strike event detection from distributed floor accelerometer signals.}
    (a) Representative raw acceleration signals $a_j(t)$ recorded simultaneously by all sensors.
    (b) Composite detection signal $g(t)$ obtained by averaging the raw accelerometer reading. Detected foot strike events $s_k$ correspond to prominent peaks that exceed a fixed threshold (orange dashed line) and satisfy a minimum separation constraint, guaranteeing one detection per step.
}
    \label{fig:TOA}
\end{figure*}

At each time instant, the absolute acceleration signals from all $N_s(=11)$ sensors are spatially averaged to form a composite detection signal,
\begin{equation}
g(t) = \frac{1}{N_s}\sum_{j=1}^{N_s} |a_j(t)| ,
\end{equation}
which can emphasize the impulsive vibration events and suppress sensor noise. The composite signal is smoothed using a short moving-average filter to attenuate high-frequency fluctuations unrelated to footstep impacts.

To detect foot strike, we define a nominal threshold as a fraction of $g(t)$'s maximal value. In this way, the footsteps can be detected when $g(t)$ rises above this threshold. In addition, to accommodate the fact that a strong foot strike can generate multiple vibrational peaks, we impose a separation principle in that consecutive peaks within 200 milliseconds belong to the same foot strike (typical adult's walking footsteps are at least 200ms apart from each other, Figure \ref{fig:TOA}). 
Via adjusting this threshold heuristically and reinforcing the separation rule, we can robustly detect the foot strike in real-time, yielding the foot strike time stamp $s_k$ for the following step.

% A nominal threshold is initialized as a fraction of the maximum value of the smoothed signal $g(t)$ and adjusted so that the detected peaks match the known number of heel-strike events while enforcing a minimum inter-peak separation corresponding to the shortest physically plausible step period. This procedure yields a robust detection threshold, which is fixed and reused during testing. During testing, footstep events are detected online using the fixed parameters learned during training. As data are streamed, prominent local maxima of $g(t)$ that exceed the threshold and satisfy the separation constraint are identified causally, yielding a detection index $s_k$ for each footstep.

%Unlike energy-based localization methods that estimate arrival windows or impact power, the detected index $s_k$ is used solely to anchor waveform extraction. As illustrated in Fig.~\ref{fig:building_reservoir}(a), a fixed-length vibration window preceding each detected peak is extracted simultaneously from all sensors and passed to the reservoir-state construction stage.

\subsubsection*{Reservoir State Vector Construction}

%\paragraph{\textbf{Vector Formalization}}

Once a foot strike is detected at $s_k$, a waveform window of $t_w$ seconds is applied immediately before $s_k$ to extract data from all sensors, yielding
\begin{equation}
\mathbf{A}_k =
\begin{bmatrix}
a_1(s_k-l+1) & \cdots & a_{N_s}(s_k-l+1)\\
\vdots & \ddots & \vdots\\
a_1(s_k) & \cdots & a_{N_s}(s_k)
\end{bmatrix}
\in \mathbb{R}^{l\times N_s},
\qquad
l = \lfloor t_w f_s \rfloor,
\label{eq:Ak}
\end{equation}
where $a_j(t)$ is the acceleration data from the $j^\text{th}$ sensor, $N_s$ is total the number of sensors, and $f_s$ is the sampling frequency. 
This multi-sensor waveform window $\mathbf{A}_k$ is then reshaped into a high-dimensional vector,
\begin{equation}
\mathbf{r}^{\mathrm{raw}}_k = \mathrm{vec}\left(\mathbf{A}_k\right)\in\mathbb{R}^{1\times d},
\; d = l N_s,
\label{eq:rraw}
\end{equation}
which serves as the raw reservoir state corresponding to the $k^\text{th}$ foot strike. This vector representation preserves the complete spatio–temporal structure of the vibration response across the sensor network. In this study, the window length $t_w$ is set 120 milliseconds to cover the dominant transient response from a single footstep while avoiding overlap with subsequent steps---yielding $d\approx1320$ for $N_s=11$ sensors sampled at $f_s\approx1$kHz.

\subsubsection*{Root Mean Square (RMS) Normalization}
Human subjects differ in their body mass, footwear, and walking gait habits; such differences introduce significant and nuisance variations in the sensor data that are not related to footstep location. To suppress these undesirable variations while preserving the spatio-temporal waveform structure that encodes the footstep location, we normalized each vectorized reservoir state by its global Root Mean Square (RMS) magnitude in that:
\begin{equation}
\mathbf{r}_k
=
\frac{\mathbf{r}_k^{\mathrm{raw}}}
{\mathrm{RMS}(\mathbf{r}_k^{\mathrm{raw}})}
=
\frac{\mathbf{r}_k^{\mathrm{raw}}}
{\sqrt{\frac{1}{d}\left\|\mathbf{r}_k^{\mathrm{raw}}\right\|_2^2}}.
\label{eq:rms_norm}
\end{equation}

\subsubsection*{Principal Component Analysis (PCA) Compression}
The normalized reservoir state vector $\mathbf{r}_k$ contains rich information; however, it also contains substantial redundancy due to correlations across sensors and time steps. To address this issue, we apply Principal Component Analysis (PCA). Suppose we want to use the reservoir state vectors from $N$ footsteps for readout weight training, we can first assemble them into a matrix $\mathbf{R}\in\mathbb{R}^{N\times d}$ so that:
\begin{equation}
    \mathbf{R} = 
    \begin{bmatrix}
        \mathbf{r}_1 \\
        \mathbf{r}_2 \\
        \vdots\\
        \mathbf{r}_N 
    \end{bmatrix}.
\end{equation}

This way, the empirical covariance matrix
\begin{equation}
    \mathbf{C}=\frac{1}{N-1}\mathbf{R}^{\top}\mathbf{R}
\end{equation}
admits an eigen-decomposition $\mathbf{C}=\mathbf{V}\mathbf{\Lambda} \mathbf{V}^{\top}$, with eigenvalues $\left\{\lambda_i\right\}$ sorted in a descending order. Then, one can project the reservoir state from step \#$k$ onto the first $D$ principal directions,
\begin{equation}
    \mathbf{z}_k = \mathbf{r}_k \mathbf{V}_D,
\end{equation}
where the matrix $\mathbf{V}_D\in\mathbb{R}^{d\times D}$ is part of $\mathbf{V}$ containing its first $D$ eigenvectors, and $\mathbf{z}_k\in\mathbb{R}^{1\times D}$ is called the first $D$ \textit{principle components} corresponding to step \#$k$. The cumulative variance of the data from all sensors is represented by the fraction of variance retained by this compression:
\begin{equation}
    \eta(D)=\frac{\sum_{i=1}^{D}\lambda_i}{\sum_{i=1}^{d}\lambda_i}.
\end{equation}

This value increases rapidly — $\eta(40)\approx97\%$ and $\eta(60)\approx99\%$ — indicating that the building’s vibration response to footstep excitation is intrinsically low-dimensional despite the raw data's high dimensionality.

\begin{figure*}[b!]
    \centering
    \includegraphics[width=0.99\textwidth]{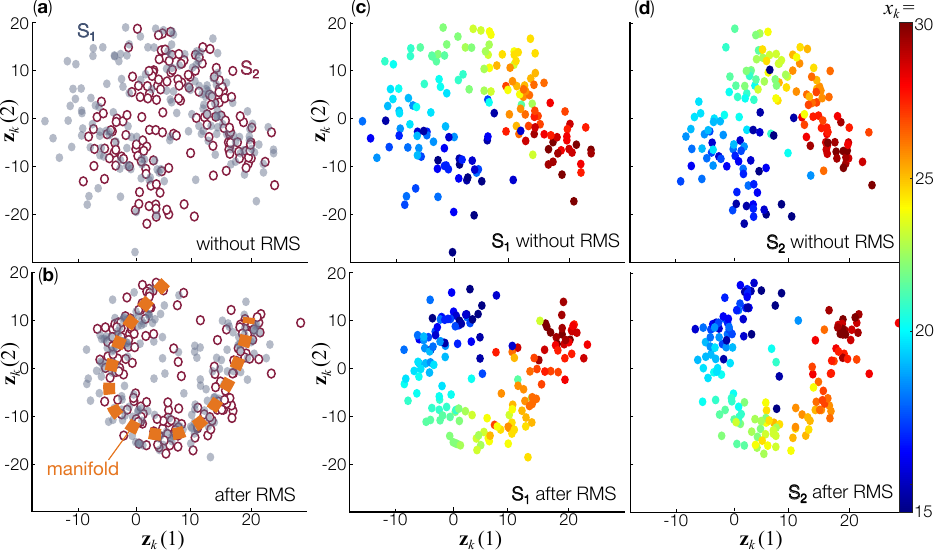}
    \caption{
       \textbf{RMS normalization aligns reservoir states across occupants and exposes location-dependent features.} In this plot, each dot corresponds to a unique footstep, based on the first two principal components of the normalized and projected reservoir state vector (i.e., $\left[\mathbf{z}_k(1), \; \mathbf{z_k}(2)\right]$, $k=1\ldots N$.
       (\textbf{a,b}) The two subjects' footstep data before and after applying RMS normalization. It's clear that after RMS, the data collapses into a compact ``manifold.''
       \textbf{(c,d)} More detailed plots, where the color maps represent the foot steps' location along the hallway (along the $x-$ axis). RMS normalization and PCA projection support physical reservoir learning across different participants. 
}
    \label{fig:rms_pca}
\end{figure*}

PCA can also help us visualize how RMS normalization enhances the features of the reservoir state space, thereby supporting PRC learning. To this end, we use the first two principal components of each foot strike, that is, the first two elements of $\mathbf{z}_k$ --- $\mathbf{z}_k(1)$ and $\mathbf{z}_k(2)$, $k=1\ldots N$ --- as the two coordinates to generate projection plots in Figure~\ref{fig:rms_pca}. In these plots, each dot represents a unique footstep. 
Figures~\ref{fig:rms_pca}(a) and ~\ref{fig:rms_pca}(b) compare the principle components \emph{before} and \emph{after} the RMS normalization, respectively. Before RMS, the results from the two subjects fall into partially separated clusters. In contrast, after RMS normalization, the two subjects' data converged on a continuous, curved low-dimensional region (often called a \emph{manifold}).  Furthermore, we color these plots according to the corresponding footstep's location along the global $x-$axis, as shown in Figures~\ref{fig:rms_pca}(c) and \ref{fig:rms_pca}(d). One can observe that after the RMS normalization, the footstep data are distributed smoothly and monotonically along the manifold's trajectory. 

%project the high-dimensional reservoir state vector $\mathbf{r}_k\in\mathbb{R}^{d}$ onto the first two principal directions of PCA and plot the resulting 2-D coordinates. As a result, \textbf{PC}$_1$ and \textbf{PC}$_2$ denote the \emph{coordinates} of a state this projection, i.e., the first two orthonormal directions in $\mathbb{R}^{d}$ that capture the largest variance in the dataset. Thus, each dot in Fig.~\ref{fig:rms_pca}(a--c) is one detected footstep, shown in the plane $(\mathrm{PC1},\mathrm{PC2})$.

Results from these PCA suggest that, before RMS normalization, the data variance captured by $\mathbf{z}_k(1)$ and $\mathbf{z}_k(2)$ are heavily influenced by \emph{amplitude-related, subject-dependent effects}, such as the subjects' body mass, footwear, impact strength, and walking gait habits. As a result, data from the same foot strike location can still be mapped to different regions in the plots. 
%
% A linear readout trained on one subject therefore learns a subject-specific affine map and generalizes poorly to the other subject. 
%
RMS normalization can remove the subject-dependent amplitude differences while preserving the relative spatio-temporal \textit{shape} of the vibration waveforms. As a result, data from the two subjects ``collapse'' and merge to the same manifold, so that the variance within this manifold are largely occupant-independent and only governed by the building’s vibratory wave propagation physics. Critically, this overlap is beneficial for the subsequent reservoir computing training because it makes the mapping from reservoir state to footstep position \emph{consistent across different subjects}. 

%Figure~\ref{fig:rms_pca}(c) confirms that the shared structure is also spatially informative: when colored by $x$-position, the embedded points vary smoothly and approximately monotonically along the trajectory in $(\mathrm{PC1},\mathrm{PC2})$. Consequently, the same linear readout can decode position for both occupants, which directly explains the reduced cross-subject error in Fig.~\ref{fig:rms_pca}(d).

\subsubsection*{Linear Readout via Ridge Regression}
After RMS normalization and PCA projection, the dynamic response from each footstep is represented by a low-dimensional reservoir state $\mathbf{z}_k\in\mathbb{R}^{1 \times D}$. 
The final step in our PRC pipeline is to map this reservoir state to a prediction of this foot strike's position using a weighted linear summation. Specifically, the estimated foot strike location $\hat{\mathbf{p}}_k=(\hat{x}_k\;\;\hat{y}_k)^\top\in\mathbb{R}^{2\times 1}$ is given by
\begin{equation}
\hat{\mathbf{p}}_k
=
\mathbf{W}_{\mathrm{out}}^\top
\begin{bmatrix}
\mathbf{z}_k\\
1
\end{bmatrix},
\label{eq:readout}
\end{equation}
where the appended constant accounts for a bias term and
$\mathbf{W}_{\text{out}}\in\mathbb{R}^{(D+1)\times 2}$ denotes the readout weight vector.
The readout weights are learned from training data using ridge regression. Denote $\mathbf{Z}_{\mathrm{tr}}\in\mathbb{R}^{N_{\mathrm{tr}}\times(D+1)}$ as the training data matrix from stacking the projected reservoir states $[\mathbf{z}_k^\top\;\;1]$ of all footsteps used for training, and matrix $\mathbf{P}_{\mathrm{tr}}\in\mathbb{R}^{N_{\mathrm{tr}}\times 2}$ collect the corresponding ground-truth locations $\mathbf{p}_k=(x_k\;\;y_k)^\top$. The readout is obtained by solving:
\begin{equation}
\mathbf{W}_{\mathrm{out}}
=
\left(\mathbf{Z}_{\mathrm{tr}}^\top \mathbf{Z}_{\mathrm{tr}}+\varepsilon \mathbf{I}\right)^{-1}
\mathbf{Z}_{\mathrm{tr}}^\top \mathbf{P}_{\mathrm{tr}},
\label{eq:ridge_closed_form}
\end{equation}
where $\varepsilon>0$ is a regularization parameter that improves numerical conditioning and mitigates overfitting. 

\subsubsection*{Optional Online Postprocessing via a 2-D Kalman Filter}
\label{sec:kf}

To reduce step-to-step jitter and smooth the outputs of physical reservoir computing, one can optionally post-process the footstep location prediction $\hat{\mathbf{p}}_k$ from Eq. (\ref{eq:readout}) using a linear Kalman filter and a constant-velocity motion model. Importantly, this step operates \emph{only} on the readout predictions and is entirely independent from reservoir construction, feature extraction, and training. Therefore, Kalman filtering does not increase the reservoir computing capacity or alter the physical representation underpinning the sensor readings. Instead, it imposes a lightweight kinematic consistency constraint across successive footsteps. 
We define the latent kinematic state of the detected footstep $\#k$ as
\begin{equation}
\mathbf{p}^*_k =
\begin{bmatrix}
x_k & y_k & \dot{x}_k & \dot{y}_k
\end{bmatrix}^\top,
\end{equation}
where $(x_k,y_k)$ is the planar footstep position and $(\dot{x}_k,\dot{y}_k)$ is the corresponding step-to-step velocities.
The state evolves according to a discrete-time, constant-velocity model, in that 
\begin{equation}
\mathbf{p}^*_{k} = \mathbf{F}\mathbf{p}^*_{k-1} + \mathbf{n}_{k-1},
\qquad
\hat{\mathbf{p}}_{k} = \mathbf{H}\mathbf{p}^*_k + \mathbf{u}_k,
\label{eq:kf_model}
\end{equation}
with state-transition and observation matrices
\begin{equation}
\mathbf{F}=
\begin{bmatrix}
1 & 0 & \Delta & 0\\
0 & 1 & 0 & \Delta\\
0 & 0 & 1 & 0\\
0 & 0 & 0 & 1
\end{bmatrix},
\qquad
\mathbf{H}=
\begin{bmatrix}
1 & 0 & 0 & 0\\
0 & 1 & 0 & 0
\end{bmatrix}.
\label{eq:kf_FH}
\end{equation}

Here, $\Delta$ is the discrete time increment between successive detected footsteps. Because the Kalman filter operates in the \emph{step index domain} rather than physical time, we set $\Delta=1$ without loss of generality. Variations in actual step timing and walking speed are instead absorbed into the process noise term: Specifically, the process noise $\mathbf{n}_k\sim\mathcal{N}(0, Q)$ captures the unmodeled variations in walking speed and direction (e.g., acceleration, deceleration, or slight turning), while the measurement noise term $\mathbf{u}_k\sim\mathcal{N}(0, R)$ describes the uncertainty in the instantaneous PRC readout predictions.
In this study, we use isotropic covariance matrices $Q=q\, \mathbf{I}_4$ and $R=r\,\mathbf{I}_2$, with scalar parameters $q$ and $r$ selected heuristically to balance the need for smoothing footstep trajectory smoothness and for preserving genuine motion changes. The filtered position estimate at each step is taken as the position component of the posterior mean,
\begin{equation}
    \hat{\mathbf{p}}^{\mathrm{KF}}_k =
\begin{bmatrix}
\hat{x}^{\mathrm{KF}}_k & \hat{y}^{\mathrm{KF}}_k
\end{bmatrix}^\top.
\end{equation}

Overall, the Kalman filtering step suppresses high-frequency, step-to-step fluctuations due to measurement noise or weak lateral observability, while preserving the global trajectory shape dictated by the PRC predictions. It operates on a four-dimensional state vector with closed-form updates, so the Kalman filter incurs negligible computational overhead and remains compatible with real-time footstep localization.

\subsubsection*{Error Metrics}
We assess the accuracy of the footsteps' prediction using Root-Mean-Square Error (RMSE) computed over all detected footsteps. For each step $k$, the overall RMSE is
\begin{equation}
    \mathrm{RMSE} = \sqrt{\frac{1}{N}\sum_{k=1}^{N} {\left\| \hat{\mathbf{p}}_k - \mathbf{p}_k \right\|_2 }^2},
\end{equation}
where $\mathbf{p}_k = [x_k\;\; y_k]^\top$ and $\hat{\mathbf{p}}_k = [\hat{x}_k\;\; \hat{y}_k]^\top$ denote the ground-truth and predicted footstep locations, respectively. And $N$ is the total number of footsteps in the evaluated data set. To assess directional localization performance, we also reported errors along the longitudinal ($x$) and lateral ($y$) directions separately:
\begin{equation}
    \mathrm{RMSE}_x = \sqrt{\frac{1}{N}\sum_{k=1}^{N} (\hat{x}_k - x_k)^2},
\qquad
\mathrm{RMSE}_y = \sqrt{\frac{1}{N}\sum_{k=1}^{N} (\hat{y}_k - y_k)^2}.
\end{equation}

\section{Results --- Successful Footstep Location Prediction Across Participants}

The experiments in this study aim to assess how effective the proposed PRC pipeline can locate the footsteps with \emph{limited training data} and \emph{limited sensors}. This reflects the central premise that the building’s nonlinear structural dynamics can perform most of the information processing. Accordingly, training and testing are organized based on walking traversals (i.e., $\text{Tr}_i$) rather than individual footsteps. Moreover, we systematically tailor the training data diversity (i.e., number of traversals) and spatial sensing coverage (i.e., number of sensors) to obtain deeper insight and best practice. 

\subsection{Single-Participant Footstep Prediction}

We first evaluate the PRC's prediction performance when the training and testing are performed using walking traversals from the same participant.
%
% This single-participant protocol evaluates whether foot strike-induced building vibrations are consistently encoded by the building into distributed accelerometer responses that can be recovered using a single linear readout across repeated traversals (Fig.~\ref{fig:singe_exp}). 
%
In this protocol, all training and testing data are from the same subject. Specifically, we select the data from several traversals $\{\text{Tr}_i\}$ for readout training (Eqs. \ref{eq:readout}-\ref{eq:ridge_closed_form}), and use the remaining traversal(s) data for testing. 

% The number of training traversals is progressively increased from one to five, and localization performance is evaluated on the unseen traversal(s). It probes how much excitation diversity—arising from natural variability in gait, impact strength, and timing across repeated walks—is required for the linear readout to generalize (Fig.~\ref{fig:singe_err}). To examine the role of spatial sensing, the same training/testing splits are repeated using randomly selected subsets of the available accelerometers. For each sensor count, multiple random subsets are evaluated, and errors are averaged, isolating how localization accuracy scales with the number of sensors (Fig.~\ref{fig:singe_err}).

\begin{figure*}[!htb]
    \centering    \includegraphics[width=0.90\textwidth]{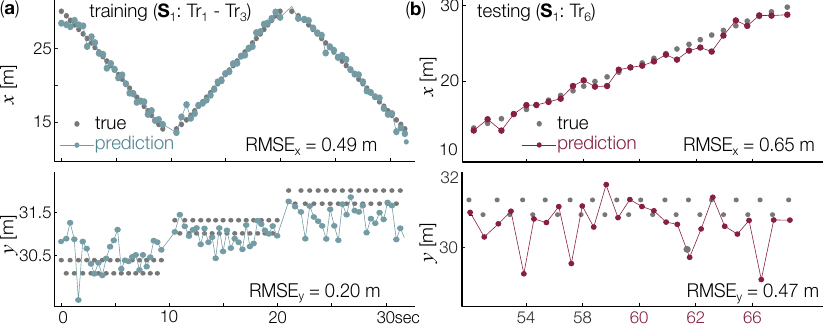}
    \caption{\textbf{The building reservoir's prediction using subject $\text{S}_1$'s data, without Kalman filtering.}
    (\textbf{a,b}) Training results using traversals $\mathrm{Tr}_1$–$\mathrm{Tr}_3$.
    (\textbf{c,d}) Testing results on the traversal $\mathrm{Tr}_6$, which is unseen during the readout training.}
    \label{fig:singe_exp}
\end{figure*}

For example, we use the normalized and projected reservoir states data from Subject 1's first three traversals (i.e., $\left\{\text{S}_1:\text{Tr}_1-\text{Tr}_3\right\}$) to train a set of linear readout, and then test this readout to the data from his sixth traversal ($\left\{\text{S}_1:\text{Tr}_6\right\}$). The accuracy of this training and testing is summarized in Figure~\ref{fig:singe_exp}. 
One can observe that, in the longitudinal direction, the building reservoir's predictions successfully follow the ground truth during both training and testing, yielding sub-meter error ($\text{RMS}=0.49$m in training and $0.65$m in testing). This result confirms that the spatio-temporal vibration dynamics generated by foot strike indeed can encode a stable, position-dependent signature along the corridor. However, footstep prediction in the lateral direction exhibits a more significant error, with RMSE increasing from $0.20$m in training to $0.47$m in testing. 
%
% This behavior reflects weaker lateral observability in a corridor geometry: small variations in foot placement, contact orientation, or boundary interaction can significantly perturb lateral vibration signatures, even for the same individual. 

\subsubsection*{Influence of Readout Training Setup}

\begin{figure*}[!ht]
    \centering
    \includegraphics[width=0.9\textwidth]{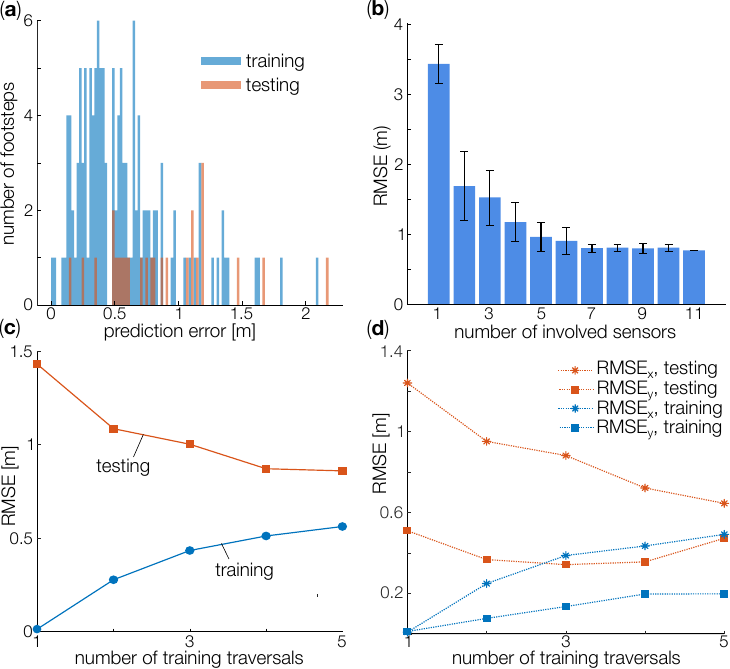}
    \caption{\textbf{Influences of physical reservoir training setup.}
    (a) Histogram of per-step Euclidean localization errors for training and testing.
    (b) Testing RMSE versus number of involved sensors.
    (c) Mean Euclidean RMSE for training and testing versus the number of traversals used for readout training.
    (d) A further breakdown of $\mathrm{RMSE}_x$ and $\mathrm{RMSE}_y$ versus number of training traversals.}
    \label{fig:singe_err}
\end{figure*}

Following the successful example in Figure~\ref{fig:singe_exp}, we further investigate the building reservoir's performance with different training setups.
First, we took a distribution-level view of prediction accuracy (Fig.~\ref{fig:singe_err}a).  Figure~\ref{fig:singe_err}(a) provides a distribution-level view of localization accuracy. Training errors are concentrated at small values, while testing errors are broader, reflecting additional variability on unseen traversals. However, the absence of extreme outliers indicates that generalization failures are gradual rather than catastrophic.
Next, we vary the number of sensors and their spatial coverage (Fig.~\ref{fig:singe_err}b). If only one floor sensor were used for reservoir computing, the RMS error is, unsurprisingly, very large. As we include additional sensors in the reservoir computing pipeline, training and testing errors decrease rapidly. Critically, footstep prediction accuracy starts to converge beyond six sensors, meaning that six floor sensors are sufficient to capture the most valuable vibrational response for information perception.
Then, we vary the number of training data (Fig.~\ref{fig:singe_err}c,d). The example shown in Fig.~\ref{fig:singe_exp} used sensor data from three traversals for readout training. Increasing the number of training traversals can reduce the testing error, but slightly increase the training error, indicating some overfitting. When the RMS errors along the longitudinal and transverse orientations are plotted separately, one can see that the benefit from additional training traversal data apply primiraly to the longitudinal component $\mathrm{RMSE}_x$ (Fig.~\ref{fig:singe_err}d). In contrast, $\mathrm{RMSE}_y$ receives a more modest improvement from additional training data, reflecting the weaker lateral observability in the floor dynamics.

% Overall, these results demonstrate that accurate single-occupant localization can be achieved with a small number of training traversals and moderate sensing coverage. The physical reservoir formed by the building dynamics produces repeatable, linearly decodable vibration states across repeated walks, enabling sub-meter localization accuracy without extensive training data or dense instrumentation.

\subsection{Cross-Participant Footstep Prediction}

Next, we evaluate whether the building reservoir pipeline can generalize across individuals with different gait characteristics. In this cross-participant setting, the linear readout $\mathbf{W}_\text{out}$ is trained using data from one subject and then tested on a second subject; meanwhile, the building, sensor layout, and signal-processing pipeline remain unchanged. This constitutes a stringent generalization test, as differences in the participants' body mass, stepping cadence, footwear, and impact conditions can significantly alter vibratory response and short-time waveform structure.

\begin{figure*}[htbp]
\centering
\includegraphics[width=0.9\textwidth]{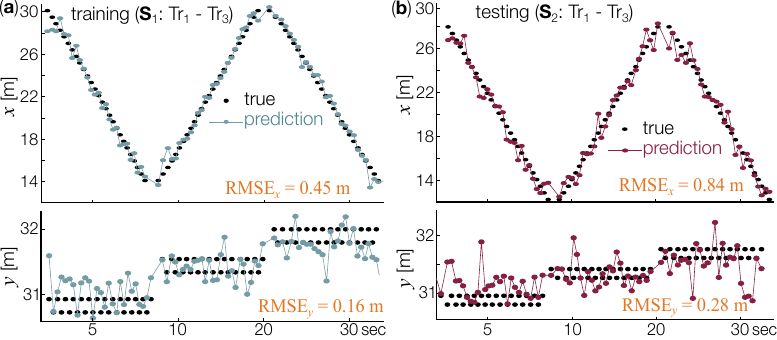}
\caption{
\textbf{Cross-occupant prediction using the building reservoir.} Training on Subject~1 (\textbf{a}) and testing on Subject~2 (\textbf{b}). The building reservoir successfully reconstructs the foot strike's trajectory across different participants.
}
\label{fig:dual_exp}
\end{figure*}

Figure~\ref{fig:dual_exp} presents an exemplary result, where the readout is trained exclusively on one participant $\left\{\text{S}_1:\text{Tr}_1-\text{Tr}_3\right\}$ and applied to all footsteps from the other participant $\left\{\text{S}_2:\text{Tr}_1-\text{Tr}_3\right\}$. Even though the building reservoir never observed the data from subject S$_2$ during training, it successfully reconstructs his footstep trajectory, correctly predicting the foot strike order and direction, without cumulative drifting. Errors are most pronounced near rapid velocity changes (i.e., turning), where step-to-step motion departs from a smooth constant-velocity trajectory. 

% More importantly, while subject-specific differences modulate vibration amplitude and introduce additional variability, the relative timing and phase relationships embedded in waveform shape remain largely dictated by the building geometry and boundary conditions, which is enabled by RMS and PCA processing on raw vibration data, allowing a single linear readout to generalize across occupants.

\subsubsection*{Effect of Kalman Filtering}
\begin{figure*}[!ht]
\centering
\includegraphics[width=0.85\textwidth]{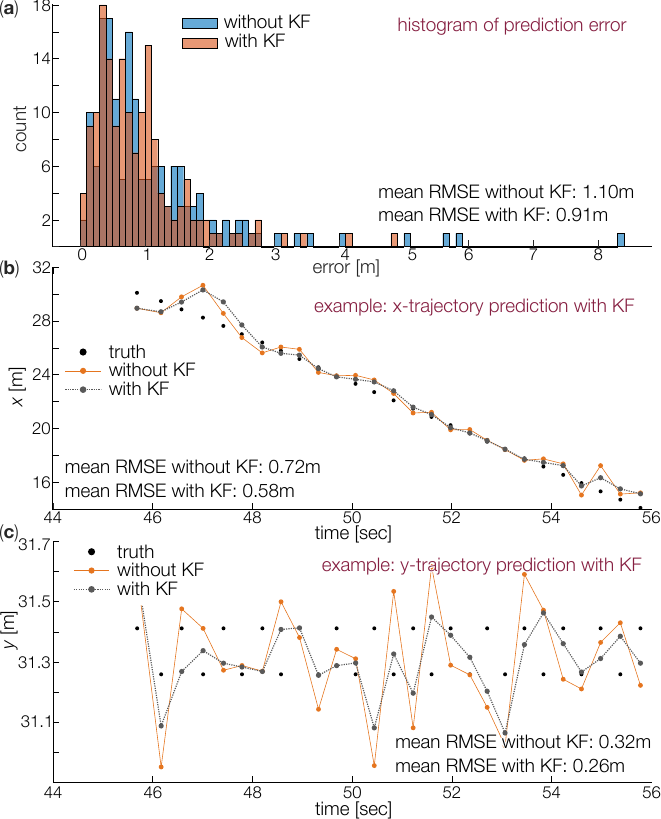}
\caption{
\textbf{Cross-participant PRC prediction with Kalman filtering.}
(\textbf{a}) Distribution of per-step prediction errors with and without Kalman filtering.
(\textbf{b}) Example trajectory along the hallway direction.
(\textbf{c}) Example trajectory across the hallway direction.
}
\label{fig:CrossOccupant_KF}
\end{figure*}

To improve temporal consistency in footstep prediction, we apply the optional, constant-velocity Kalman filter described in Section~\ref{sec:kf} as a post-processing step. Figure~\ref{fig:CrossOccupant_KF} compares PRC predictions with their Kalman-filtered counterparts.
Kalman filtering primarily narrows the central portion of the error distribution and reduces moderate errors, lowering the mean RMSE from $1.10$m to $0.91$m. In particular, large outliers are reduced --- this is expected because filtering can enforce short-term kinematic consistency. The example trajectories in Fig.~\ref{fig:CrossOccupant_KF}(b,c) illustrate this effect: high-frequency step-to-step fluctuations are suppressed, yielding smoother and more physically plausible trajectories. Importantly, Kalman filtering refines but does not fundamentally alter the underlying PRC predictions.

\subsubsection*{Influence of Readout Training Setup}
\begin{figure*}[!ht]
\centering
\includegraphics[width=0.9\textwidth]{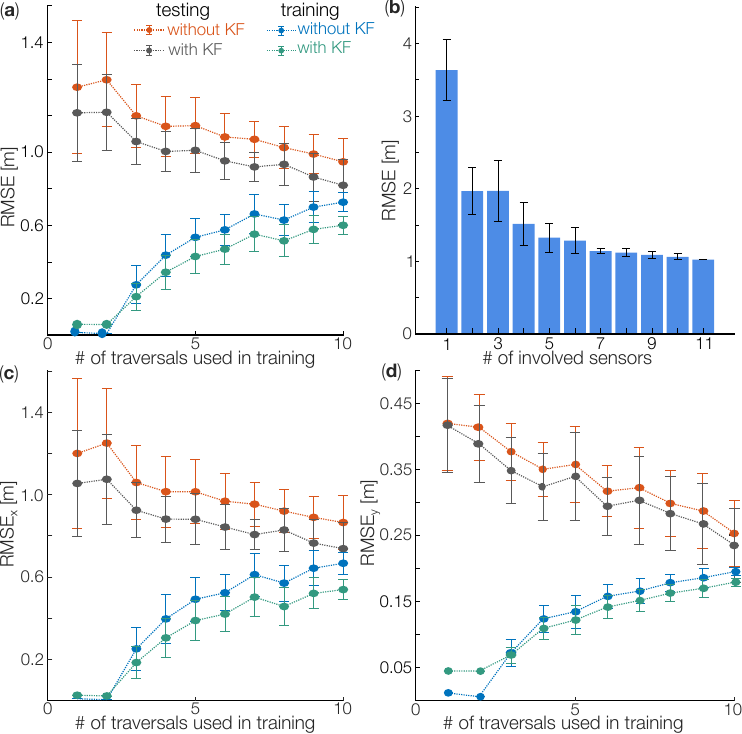}
\caption{
\textbf{Influence of training data setup and sensor selections on cross-occupant localization accuracy.}
(a) Training and testing RMSE versus number of training traversals.
(b) Cross-occupant test RMSE versus number of sensors.
(c,d) Further breakdown of testing and testing errors along the two orientations, $\mathrm{RMSE}_x$ and $\mathrm{RMSE}_y$.
}
\label{fig:paths_error}
\end{figure*}

Similar to the single-participant test, we quantify how varying the readout training setup affects cross-occupant generalization. We first vary the number of traversals used in readout training. More specifically, we construct training data sets by randomly selecting between one and ten traversals from the combined pool of twelve traversals (six per subject). Meanwhile, the remaining traversals are used for testing. Results are summarized in Figure~\ref{fig:paths_error}.

Figures~\ref{fig:paths_error}(a,c,d) show that cross-participant generalization improves steadily as more traversal data are employed in training. With only one or two training traversals, testing errors are large and highly variable. As additional traversals are incorporated into readout training, testing RMS error decreases in both magnitude and variation, demonstrating that broader excitation diversity enables the reservoir representation to better capture footstep dynamics across different participants. Performance stabilizes after approximately six to eight training traversals, beyond which further improvements are marginal. 
It is worth noting that the breakdown of RMS error in Fig.~\ref{fig:paths_error}(d) reveals that these improvements apply primarily to the longitudinal coordinate $\mathrm{RMSE}_x$, similar to those in the single-participant test. Also worth noting is that Kalman filtering consistently reduces absolute error but does not fundamentally change the overall trend. 

% In contrast, $\mathrm{RMSE}_y$ exhibits weaker and more gradual improvement, reflecting the lower lateral observability of the corridor and its greater sensitivity to subject-specific walking patterns. Once sufficient excitation diversity is achieved, both components converge, indicating that the remaining error is dominated by structural, rather than data-limited, effects.

Next, we vary the number of sensors and their spatial coverage (Fig.~\ref{fig:paths_error}b). Similar to the single-participant test, as more sensors are included in the readout training, test RMS error decreases rapidly, reaching approximately $1.5$m with four sensors and converging to $1.0$--$1.1$m beyond six sensors. % This saturation mirrors the training-path results and highlights the complementary roles of excitation diversity and spatial sensing coverage in stabilizing cross-subject inference.

% Across all panels, Kalman filtering consistently reduces absolute error but does not change the dependence on training-traversal or sensor count. This confirms that cross-occupant generalization is governed primarily by the quality of the physical reservoir encoding, rather than by post hoc temporal smoothing.

Overall, the cross-participant test results demonstrate that the vibration-based building reservoir computing can provide a mostly subject-agnostic representation of footstep-induced dynamics. With moderate sensing coverage and limited training data, the system achieves stable, meter-scale prediction accuracy without requiring extensive instrumentation or subject-specific retraining.

\section{Discussion}

%This section interprets the experimental results from the perspective of PRC, with an emphasis on what spatial information is mechanically encoded by the building–sensor system and how this encoding constrains achievable localization performance. 

% \subsection{Overall Evaluation of Vibration-Based PRC Localization Performance}

Taken together, the test results demonstrate that treating the instrumented building as a physical reservoir provides us with a simple yet powerful tool for locating footsteps. 
%
% Beyond achieving relative low average localization error, 
%
The proposed physical reservoir computing pipeline preserves the \emph{geometric and temporal structure} of walking trajectories --- the predicted footsteps follow the participants' traversals along and cross the hallway (without collapsing toward the hallway's centerline), maintain correct footstep order, and reproduce the natural variations in walking speed and lateral motion --- all with a low average error.
The reservoir outcomes are consistent across training and testing, and remain stable under cross-participant tests. 
Critically, these advantages are achieved using only short windows of waveform data, simple data processing (RMS and PCA), and a linear readout, indicating that the performance gains arise from the physical spatio-temporal encoding provided by the building structure itself, rather than from increased model complexity, recursive filtering, or trajectory-level constraints.

\subsection{Comparison with Energy-Based Methods}

Here, we quantitatively compare the performance of our reservoir computing pipeline with that of prior work using energy-based methods on the same dataset (Table~\ref{tab:rmse_compare}).
These energy methods include heuristic Received Signal Strength (RSS) approaches, modified heuristic variants, and maximum-likelihood (MaxLik) estimators. They all used identical sensor layouts and test data, but fundamentally differ in formulation: energy-based methods rely on handcrafted energy features and attenuation models, whereas PRC operates directly on high-dimensional vibration waveforms using a learned linear readout.
We observe two significant advantages of our proposed PRC pipeline over the published energy-based methods. 

(1) \textit{Reduced Error}: 
The building reservoir pipeline with RMS normalization and PCA compression achieved the lowest overall Euclidean RMSE across both test subjects. Relative to the heuristic RSS baseline, PRC reduces total RMSE by approximately $38\%$ for Subject~1 and $33\%$ for Subject~2. 

(2) \textit{Cross-Participant Prediction}: 
It is important to emphasize that most existing vibration-based footstep studies—including the energy-based methods listed here—primarily report results \textit{on the same participant}. That is, training and testing data were drawn from the same individual, or subject-dependent parameters (e.g., attenuation coefficients or reference energy profiles) are implicitly calibrated using all available data. Explicit cross-participant evaluation, where the prediction model is trained on one individual and then tested on a different individual, is rarely reported.  There is no direct evidence on where the energy method can achieve cross-participant prediction; nonetheless, our results highlighted in Table~\ref{tab:rmse_compare} present a promising new direction.

%(3) \textit{Prediction Footstep Location In the Transverse Direction?}: 
% 
% In contrast, the PRC framework is explicitly evaluated under both single-occupant and cross-occupant conditions. The results in Table~\ref{tab:rmse_compare} therefore reflect a more stringent setting: for $S_2$, the PRC readout is trained without access to that subject’s data, whereas the reported energy-based results rely on subject-specific calibration inherent to their formulation. As such, the comparison is not intended as a strictly fair machine-learning benchmark, but rather as a \emph{contextual reference} that highlights how much spatial information can be accessed when different physical features are extracted from the same structure, sensor layout, and dataset.

\begin{table*}[htb]
\centering
\caption{RMSE comparison for foot strike location prediction. The units are in meters. Note that the starred ($^*$) values are cross-participant test results, and none of the prior studies investigated this.}
\small
\label{tab:rmse_compare}
\begin{tabular}{lcccccc}
\toprule
\multirow{2}{*}{Algorithm} 
 & \multicolumn{3}{c}{Test Subject 1} 
 & \multicolumn{3}{c}{Test Subject 2} \\
 & RMSE$_x$ & RMSE$_y$ & RMSE$_T$
 & RMSE$_x$ & RMSE$_y$ & RMSE$_T$ \\
\midrule
Heuristic RSS~\cite{alajlouni2019new}
 & 1.39 & 0.35 & 1.43
 & 1.36 & 0.35 & 1.40 \\

Modified Heuristic~\cite{alajlouni2020passive}
 & 0.79 & 0.66 & 1.03
 & 0.93 & 0.64 & 1.13 \\

MaxLik RSS~\cite{alajlouni2022maximum}
 & 0.67 & 0.58 & 0.89
 & 0.74 & 0.58 & 0.94 \\

\midrule
Traditional PRC (no RMS, no PCA)
 & 1.21 & 0.34 & 1.33
 & 2.60$^*$ & 0.37$^*$ & 2.69$^*$ \\

Vibration-based PRC (This work)
 & \textbf{0.63} & \textbf{0.47} & \textbf{0.86}
 & \textbf{0.98}$^*$ & \textbf{0.32}$^*$ & \textbf{1.13}$^*$ \\

\bottomrule
\end{tabular}
\end{table*}

\subsection{Why Is Predicting Foot Strike's Lateral Location More Challenging?}

\begin{figure*}[!hbt]
 \centering
 \includegraphics[width=0.8\textwidth]{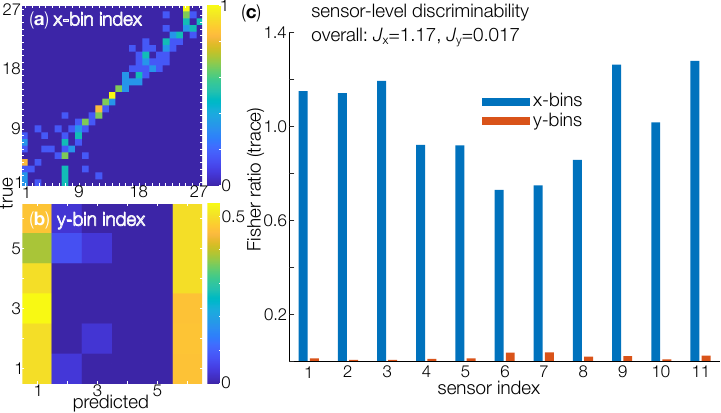}
 \caption{
  \textbf{Observability of footstep positions in longitudinal and lateral orientations.}
  (a) Confusion matrix for discretized $x$-bin location.
  (b) Confusion matrix for discretized $y$-bin location.
  (c) Sensor-level Fisher ratio for $x$-bins and $y$-bins computed directly from raw vibration features.}
  \label{fig:confusion_xy}
\end{figure*}

Figures~\ref{fig:confusion_xy}(a) and (b) show the confusion matrices obtained by discretizing the hallway space into spatial ``bins'' and comparing the true bin of each footstep with the PRC prediction. A strong diagonal pattern in the confusion matrix indicates that spatial locations of foot strikes are clearly distinguishable, whereas an off-diagonal spreading indicates ambiguity.
As shown in Figure~\ref{fig:confusion_xy}(a), the confusion matrix along the hallway (i.e., $x-$ direction) is sharply diagonal, with errors largely confined locally. This indicates strong physical separability in this direction. In contrast, the $y$-direction confusion matrix shown in Figure~\ref{fig:confusion_xy}(b) shows pronounced off-diagonal spread and partial collapse into a small number of bins, indicating that footsteps at different lateral positions often produce very similar vibration responses and are therefore difficult to distinguish by the PRC pipeline. 
 %
 %This anisotropy arises from the physics of wave propagation in corridor-like structures. Longitudinal position primarily modulates wave travel distance, arrival timing, and phase alignment across the sensor network, producing smooth and repeatable spatio-temporal patterns. Lateral position, by comparison, is encoded mainly through subtle energy imbalance and reflection effects, which are more sensitive to gait variability, foot orientation, and local boundary interactions. As a result, different $y$ locations frequently excite overlapping reservoir states, leading to intrinsic ambiguity that cannot be resolved by the readout alone.

Figure~\ref{fig:confusion_xy}(c) provides complementary evidence directly at the sensor level using the Fisher discriminability analysis ~\cite{giorgetti2008localization}. For each accelerometer $j$, a Fisher ratio can be computed as
\begin{equation}
J_j =
\frac{\sum_{c} N_c (\mu_c - \mu)^2}
     {\sum_{c} N_c \sigma_c^2},
\label{eq:fisher_ratio}
\end{equation}
where each spatial bin $c$ corresponds to a group of footsteps at similar locations, $\mu_c$ and $\sigma_c^2$ are the mean and variance of the sensor’s vibration feature within that bin, $N_c$ is the number of footsteps in the bin, and $\mu$ is the global mean across all bins. \emph{Physically, the numerator of Eq. (\ref{eq:fisher_ratio}) measures how much a sensor reading changes with respect to foot strike position, while the denominator measures how much that response fluctuates due to noise, gait variability, and other non-critical effects}. Therefore, a large Fisher ratio indicates that a sensor reading is sensitive to the changes in foot strike locations, and vice versa. 
To this end, results in Figure~\ref{fig:confusion_xy}(c) show that, across all sensors, the Fisher ratios are consistently high along the $x-$ direction and close to zero for $y$, indicating that the foot strikes' longitudinal positions are already linearly separable based on the raw vibration measurements.

Therefore, from the fundamental physics point of view, the confusion matrix and Fisher discriminability analysis conclude that the vibratory response of the building structure can inherently distinguish foot strike location along the $x$-axis --- so that RPC pipeline can extract this information accurately with a simple linear readout --- but the information about the $y-$axis location is compressed. Consequently, \emph{improving the prediction accuracy along the $y-$axis will require physically reconfiguring the building reservoir}, such as redistributing the floor sensors, introducing new sensors with additional sensing modalities, or adjusting the building construction setup to further enrich the nonlinear dynamics.

\section{Conclusion}

This study explores a new approach to locating building occupants’ footstep location by \emph{re-purposing an instrumented building as a physical reservoir}, whose intrinsic structural dynamics can perform machine learning in the physical domain. Different from physical-based modeling or purely data-driven machine learning, the proposed physical reservoir computing (PRC) approach exploits the physical phenomena that the building’s nonlinear spatio–temporal response can transform footstep excitations into feature-rich vibration waveforms, which inherently encodes the information about foot strike location. Therefore, a simple linear readout of these waveforms can extract this information.

A central contribution of this study is a \emph{novel vibration-based PRC pipeline that combines short-waveform reservoir states with RMS normalization and PCA projection}. RMS normalization suppresses subject-dependent variability in the sensing data, while PCA projection selects the dominant vibration modes that are most sensitive to foot strike location. Together, these data processing techniques substantially improve the accuracy of the building reservoir, enabling accurate cross-participant prediction without subject-specific retraining.

\emph{By leveraging nonlinear structural dynamics as the primary mechanism for feature expansion, the PRC pipeline also minimizes training data requirements.} Reliable footstep location prediction is achieved using only a small number of training traversals and a moderate number of sensors, in contrast to conventional machine-learning approaches that typically require large labeled datasets and extensive retraining. The footstep prediction performance converges quickly as training data coverage and sensor count increase, highlighting the data-efficiency of the physical reservoir paradigm.

\emph{The proposed method also demonstrates competitive, and often superior, performance compared to traditional energy-based footstep locating methods, particularly near the boundaries of the instrumented floor space}. This improvement underscores the value of exploiting full waveform structure rather than amplitude-only features. Finally, the confusion matrix and Fisher ratio analyses reveal that the performance gap between longitudinal and lateral direction arises from the physical limits of the building reservoir, rather than algorithmic shortcomings. 

Overall, this study establishes vibration-based physical reservoir computing as a scalable, data-efficient, and real-time framework for intelligent sensing, with broad applicability to buildings, bridges, and other engineered systems where distributed vibrations naturally encode task-relevant information.

\medskip
\textbf{Acknowledgements} 
Funding for this work was provided by NSF DCSD \#2328522, and Virginia Tech (via startup funding) 

\medskip

\bibliographystyle{elsarticle-num}
\bibliography{ref}

\end{document}